%% file: main.tex
\documentclass[acmsmall]{acmart}

\usepackage{flushend}
\usepackage[english]{babel}
\usepackage[latin1]{inputenc}
\usepackage{mathrsfs}
\usepackage{graphicx}

\usepackage{amssymb}
\usepackage{amsfonts}
\usepackage{url}
\usepackage{caption}
\usepackage{multirow}
\usepackage{longtable}
\usepackage{rotating}
\usepackage{multirow}
\usepackage{mathrsfs}
\usepackage{subfigure}
\usepackage{enumitem}
\usepackage[linesnumbered,algoruled,boxed,lined]{algorithm2e}
\usepackage{adjustbox}
\usepackage{hyperref}
\usepackage{arydshln}

\newcommand{\tabincell}[2]{\begin{tabular}{@{}#1@{}}#2\end{tabular}}

\usepackage{pgfplots}
\usetikzlibrary{pgfplots.dateplot}

\usepackage{filecontents}
\definecolor{tblue}{RGB}{31,119,180}
\definecolor{torange}{RGB}{255,127,14}
\definecolor{tgreen}{RGB}{44,160,44}
\definecolor{tred}{RGB}{214,39,40}
\definecolor{tpurple}{RGB}{148,103,189}

\newcommand{\tp}{^{\top}}
\newcommand{\inv}{^{-1}}
\newcommand{\eg}{{\it e.g.}}
\newcommand{\ie}{{\it i.e.}}
\newcommand{\etal}{{\it et al.}}
\newcommand{\wrt}{\textit{w}.\textit{r}.\textit{t}}

\newcommand{\hide}[1]{} 

\AtBeginDocument{%
  \providecommand\BibTeX{{%
    \normalfont B\kern-0.5em{\scshape i\kern-0.25em b}\kern-0.8em\TeX}}}

\setcopyright{acmlicensed}
\acmJournal{TOIS}
\acmYear{2021} \acmVolume{1} \acmNumber{1} \acmArticle{1}
\acmMonth{1} \acmPrice{15.00}\acmDOI{10.1145/3467023}

\acmJournal{JACM}
\acmVolume{37}
\acmNumber{4}
\acmArticle{111}
\acmMonth{8}

\def\model{CRANet}

\begin{document}

\title{Collaborative Reflection-Augmented Autoencoder Network for Recommender Systems}






\author{Lianghao Xia}
\affiliation{%
  \institution{South China University of Technology}
   \city{Guangzhou}
  \country{China}}
  \email{cslianghao.xia@mail.scut.edu.cn}

\author{Chao Huang}
\affiliation{%
  \institution{University of Hong Kong}
   \city{Hong Kong}
  \country{Hong Kong}}
  \email{chaohuang75@gmail.com}

\author{Yong Xu}
\authornote{Yong Xu is the corresponding author.}
\affiliation{%
  \institution{South China University of Technology}
   \city{Guangzhou}
  \country{China}}
  \email{yxu@mail.scut.edu.cn}
  
\author{Huance Xu}
\affiliation{%
  \institution{South China University of Technology}
   \city{Guangzhou}
  \country{China}}
  \email{cshuance.xu@mail.scut.edu.cn}
  
\author{Xiang Li}
\affiliation{%
  \institution{Australian National University}
   \city{Canberra}
  \country{Australia}}
  \email{u6498493@anu.edu.au}

\author{Weiguo Zhang}
\affiliation{%
  \institution{South China University of Technology}
   \city{Guangzhou}
  \country{China}}
  \email{wgzhang@scut.edu.cn}






\begin{abstract}
As the deep learning techniques have expanded to real-world recommendation tasks, many deep neural network based Collaborative Filtering (CF) models have been developed to project user-item interactions into latent feature space, based on various neural architectures, such as multi-layer perceptron, auto-encoder and graph neural networks. However, the majority of existing collaborative filtering systems are not well designed to handle missing data. Particularly, in order to inject the negative signals in the training phase, these solutions largely rely on negative sampling from unobserved user-item interactions and simply treating them as negative instances, which brings the recommendation performance degradation. To address the issues, we develop a \underline{C}ollaborative \underline{R}eflection-Augmented \underline{A}utoencoder \underline{N}etwork (\model), that is capable of exploring transferable knowledge from observed and unobserved user-item interactions. The network architecture of \model\ is formed of an integrative structure with a reflective receptor network and an information fusion autoencoder module, which endows our recommendation framework with the ability of encoding implicit user's pairwise preference on both interacted and non-interacted items. Additionally, a parametric regularization-based tied-weight scheme is designed to perform robust joint training of the two-stage \model\ model. We finally experimentally validate \model\ on four diverse benchmark datasets corresponding to two recommendation tasks, to show that debiasing the negative signals of user-item interactions improves the performance as compared to various state-of-the-art recommendation techniques. Our source code is available at https://github.com/akaxlh/CRANet.
\end{abstract}


\maketitle

\input{intro}
\input{relate}
\input{solution}
\input{eval}

\input{conclusion}

\bibliographystyle{ACM-Reference-Format}
\bibliography{refs}

\appendix

\end{document}

%% file: intro.tex
\section{Introduction}
\label{sec:intro}

The past decades have witnessed the explosive growth of online services, \eg, social media, e-commerce and online advertising, which has provided overwhelming choices for users~\cite{2019online,ren2019deep,zhao2019cross}. To accurately characterize users' preferences and identify their interested items, personalized recommendation has become an important component in many customer-oriented applications to predict user behavior~\cite{he2018nais,lian2020product}. Among various recommendation methods, Collaborative Filtering (CF) has become a key technique to infer the user's interests from not only his/her past interaction behavior, but also the behavior from other individuals~\cite{ekstrand2011collaborative,chen2019joint}. At the core of collaborative filtering methods is how to parameterize users and items with effective feature vectors based on their observed historical interactions~\cite{xue2019deep}.

With the advent of deep learning techniques, significant efforts have been made to develop neural collaborative filtering models and show performance superiority over conventional matrix factorization techniques~\cite{zhang2016discrete}. For example, Multi-layer perceptron (MLP) has been utilized to capture non-linear user-item interactions~\cite{he2017neural}. Auto-encoder architecture has served as an effective solution for dimensionality reduction by learning underlying patterns from the interaction data between users and items~\cite{strub2015collaborative, strub2016hybrid, du2018collaborative}. Furthermore, several graph neural network models aim to model relational information over the user-item interaction graph via different aggregation functions, such as graph convolution operation~\cite{zhang2019star,wang2019neural}, and attention-based message passing frameworks~\cite{song2019session,xia2021knowledge}.

To address the above limitation, this work proposes to learn complex user's preference from not only his/her positive feedback (\eg, click or visit), but also the knowledge of his/her negative and unobserved interactions with items in a explicit way. However, there exist several key technical challenges that remain to be solved for realizing the high potential predictability of user's preference with the collectively modeling of positive and unobserved user-item interactions. \emph{First}, different types (\ie, positive, negative and unobserved) of user-item interactions interweave with each other in complex ways, how to effectively encode cross-modal collaborative signals, remains a significant challenge. \emph{Second}, due to the fact that the negative and unobserved instances dominate the user-item interactions, it is crucial to alleviate such overfitting phenomenon when encoding the different-dimensional relational structures between different users and items. \emph{Third}, the sparsity degrees of interaction data vary by users, performing robust representation learning on different users and items is a key factor for effective recommendations.

With motivations and challenges above, this paper develops a \underline{C}ollaborative \underline{R}eflection-Augmented \underline{A}utoencoder \underline{N}etwork (\model) to improve the performance of collaborative filtering for recommendations, with the exploration of the observed and unobserved user-item interactions. The proposed \model\ captures the implicit users' preference from their unobserved interactions over items with two main model designs: i) We first design an integrative neural architecture with a reflective reception network and a tired-weight gated scheme, to encode the preference of users over their non-interacted users based on backward embedding propagation paradigm. With the reflection-based imputation scheme, we inject the collaborative signals from unobserved interactions into the user representation phase. ii) Additionally, our information fusion framework is built upon the autoencoder model to capture the latent dependency between observed positive user-item interactions and unobserved ones. With the reconstruction-guided learning objective, the interaction dependent structures are encoded through the stacked multiple projection layers in the modeling of collaborative effects. With the cooperation of the reflection receptor network and an autoencoder architecture, we promote the collaboration of cross-modal user-item interactions to characterize user's interests. In addition, a regularization-based tied-weight scheme is introduced to yield a robust user-item interaction embedding learning under data imbalance, and offer efficient training for the autoencoder-based reflection network.



The main contributions of this work are listed as follows.
\begin{itemize}[leftmargin=*]

\item We highlight the benefits of exploiting cross-modal user-item interactions, to allow the effectively joint modeling of both observed and unobserved interactive behavior of users.\\\vspace{-0.1in}

\item We propose a two-stage solution, \model, integrating a simple but effective reflective receptor network with an information fusion autoencoder architecture, which encodes the collaborative signal in the form of observed and unobserved interactions.\\\vspace{-0.1in}

\item Furthermore, another contribution is the effective design of a regularization-based tied weight scheme in our \model\ framework, which can alleviate the overfitting issue and offer robust embedding learning over user-item interactions by correcting the sparsity bias. A theoretical analysis is provided to show the rationality of our designed hierarchically structured gate-based reflective receptor framework in capturing user-item relations.\\\vspace{-0.1in}

\item We perform extensive experiments on several publicly accessible datasets corresponding to two recommendation tasks (\ie, rating prediction and item recommendation), and compare our \model\ framework with various state-of-the-art baselines. The model ablation study verifies the efficacy of our developed modules, and the utility of reflection mechanism in encoding implicit user preferences. In addition, we also investigate the model scalability and compare \model\ with state-of-the-art recommender systems, to show its efficiency.

\end{itemize}

The rest of this paper is organized as follows. We discuss the related work in Section~\ref{sec:relate}. The technical details of our developed \model\ framework in Section~\ref{sec:solution}. We also perform theoretical analysis of our model. Section~\ref{sec:eval} presents the evaluation results of our \model\ method as compared to state-of-the-art baselines. We finally conclude this work in Section~\ref{sec:conclusion}.

%% file: relate.tex
\section{Related Work}
\label{sec:relate}

\subsection{Neural Network-based Recommender Systems}
Many efforts have focused on proposing recommendation models with different types of neural networks~\cite{he2020lightgcn,chen2020try,chen2020block,xia2020multiplex}, such as neural network-augmented recommendation via stacking several feed-forward layers~\cite{he2018outer,he2017neural} and neural auto-regressive learning frameworks~\cite{zheng2016neural,du2018collaborative}. More recently, several graph neural networks (\eg, graph convolutional network) have been applied to recommendation tasks to model the graph-structured relations between users and items~\cite{wu2019session,zhang2019star,wang2019neural}. Different from these state-of-the-art solutions, this work aims to promote the cooperation among user-item relations and significantly improves the recommendation performance in terms of both effectiveness and efficiency, with the cross-modal reflective autoencoder network and efficient reflection mechanism.

\subsection{Relation-aware Recommendation Models}
There exist many research work aiming to investigate various relationships among users and items for recommendation applications. For example, for user side, social relations between users have been considered as the side information to improve the recommendation performance~\cite{fan2019graph,wang2019social,social2021knowledge}. For item side, knowledge graph serve as another key source to model relationships across different items~\cite{wang2019knowledge,fu2020fairness}. In addition, multiple explicit item relation modeling~\cite{xin2019relational} and user behavior learning~\cite{multibehavior2021} have been integrated with the collaborative filtering with graph neural networks. Different from these work focusing on exploring explicit relations between users and items, this work explores the implicit user preference from unobserved user-item relations without any auxiliary information. 

\subsection{Autoencoder Collaborative Filtering Framework}
Autoencoder has emerged as an important architecture to enable the collaborative filtering techniques to map the explicit user-item interactions into latent low-dimensional representations, with the objective of reconstructing the original inputs~\cite{ouyang2014autoencoder}. For instance, AutoRec~\cite{sedhain2015autorec} is the first AE-based collaborative filtering method that shows noticeably superior performance over the conventional recommendation models. Motivated by this work, many extensions have been developed to solve the collaborative filtering challenges (\eg, non-linear latent-variable modeling or data sparsity) with various autoencoder variants, such as variational AE~\cite{liang2018variational,wu2019one}, contractive AE~\cite{zhang2017autosvd} and denoising AE~\cite{wu2016collaborative}. However, most of these autoencoder modules require sufficient user-item interactions to learn accurate user/item embeddings. To fill this gap, our \model\ enables a practical learning scenario to capture implicit user-item relations with a designed cross-modal autoencoder network.

\subsection{Imputation-based Recommendation Methods}
Another research line relevant to this work lies in the recommendation methods based on the data imputation techniques. For example, the early study S-Imp~\cite{mazumder2010spectral} proposes a regularized low-rank SVD method to replace the missing elements with derived values. With the advance of deep learning techniques, Lee~\etal~\cite{lee2018impute} develops a neural network-based approach to use the inferred imputed values for learning user preference. Wan~\etal~\cite{wan2019sparse} designs an ensemble learning to perform the continuous imputation and optimize the neural matrix factorization framework. Furthermore, a zero imputation algorithm is introduced to handle missing data with sparsity normalization~\cite{yi2019not}. Along with this line and motivated by these work, we develop a reflection-augmented autoencoder network to capture implicit user preference with an effective imputation mechanism.

%% file: solution.tex
\section{Methodology}
\label{sec:solution}

\begin{figure*}
    \centering
    \includegraphics[width=0.95\textwidth]{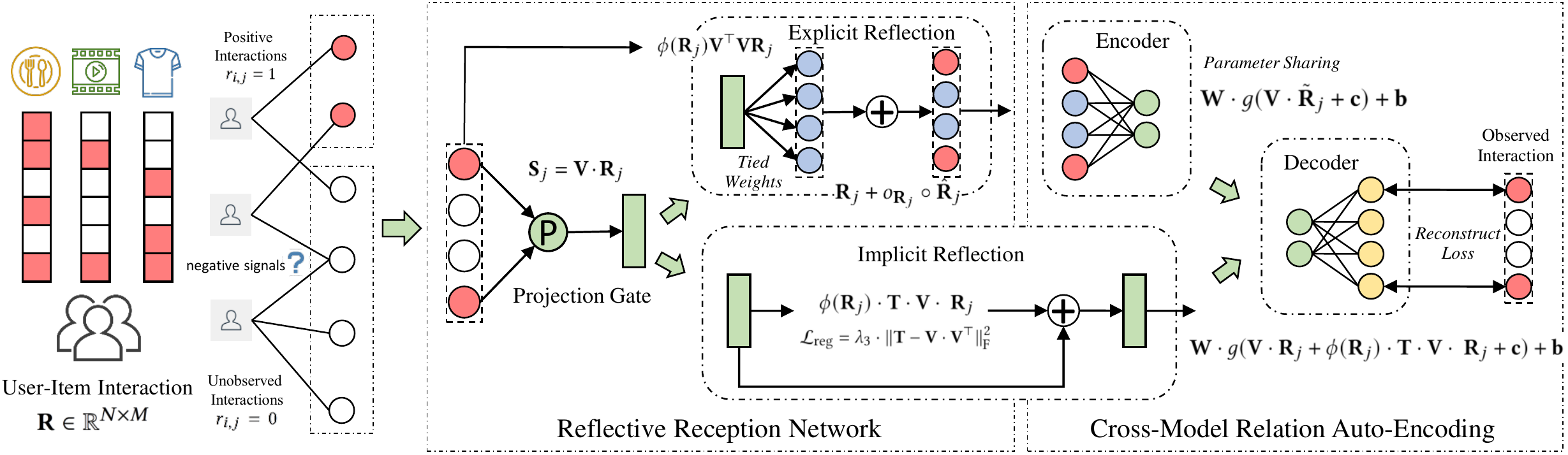}
    \vspace{-0.05in}
    \caption{Model architecture of the proposed \model. Explicit reflective reception network (\ie~utilizing $\textbf{V}^\top$ to infer the missing interaction scores as defined by Eq~\ref{equation:inference} $\sim$ Eq~\ref{equation:aepred}), and implicit reflection (\ie~employing $\textbf{T}$ to replace $\textbf{V}^{\top}$ in the reflection and the encoder in the cross-model auto-encoding as defined by Eq~\ref{equation:pred} $\sim$ Eq~\ref{equation:refreg}).}
    \vspace{-0.05in}
    \label{fig:framework}
\end{figure*}

In this section, we describe the technical details of our \model\ framework (the model architecture is shown in Figure~\ref{fig:framework}). In particular, we first introduce key definitions and preliminaries. There are two key modules in the \model\ framework corresponding to our motivations and challenges: (1) the reflective reception network that captures users' implicit preference on their non-interacted items, with the backward embedding propagation paradigm; ii) a collaborative dependency auto-encoding component which aggregates the information from different user-item relational signals. Finally, we offer a theoretical analysis for our designed encoding function of complex relations between user and item, and discuss the model complexity.

\subsection{Preliminaries}
We consider a scenario with $M$ items and $N$ users. Their interaction matrix is denoted by $\textbf{R} \in\mathbb{R}^{N\times M}$, where the corresponding interaction (\eg, explicit user rating) between the user $u_i$ ($1\leq i \leq N$) and item $t_j$ ($1\leq j \leq M$) is denoted as $r_{i, j}$ ($r_{i,j}$ is set as 0 if there is no observed interaction). In the practical recommendation scenario, the data scarcity of user-item interactions poses challenges in learning accurate user preferences. However, the interaction value $r_{i,j}=0$ does not necessarily indicate that user $u_i$ does not like item $t_j$, it is possible that the user is not aware of that item. With the observed interaction matrix $\textbf{R}$, the objective of this work aims at learning user's interests over items, the result of our learning framework is expected to predict the unknown values in $\textbf{R}$, with the exploration of observed and unobserved interactive relations across users and items to address the sparsity bias challenge. 

\begin{table*}
	\centering
	\scriptsize
	\vspace{-0.1in}
	\caption{Summary of key notations.}
	\vspace{-0.15in}
	\label{tab:dataset}
	\setlength{\tabcolsep}{1mm}
	\begin{tabular}{cc}
		\toprule
		Notation & Description \\
		$u_i$ & the $i$-th user ($1\leq i \leq M$)\\
		$t_j$ & the $j$-th item ($1\leq j \leq N$)\\
		$\textbf{R}_{i}\in\mathbb{R}^{N}$ & the interaction vector of user $u_i$ over different items \\
		$\textbf{R}_{j}\in\mathbb{R}^{M}$ & the interaction vector of user $t_j$ with different user \\
		$\alpha$ & hyperparameter which controls the decay strength \\
		$\phi(\cdot)$ & decay function which is reflective of energy dissipation in our reflection gate\\
		$o_{\textbf{R}_{j}}$, $\circ$ & zero indicator and element-wise product in our reflection gate\\
		$\textbf{V}$, $\textbf{V}^\top$ & the projection weight matrix and its corresponding transpose form for reflective gating mechanism\\
		$\tilde{\textbf{R}}$ & the recalibrated relation-aware user-item interactions\\
		$\textbf{U}\in\mathbb{R}^{N\times d_p}$ & learnable transformation matrix in our tied-weight scheme for training efficiency\\
		$\textbf{W},\textbf{V},\textbf{c},\textbf{b}$ & transformatino matrices and bias terms of user-item relation auto-encoding\\
		$\mathcal{L}_\text{reg}$ & regularization term which simulates the reflection process in encoding the implicit user-item relations\\
		\hline
	\end{tabular}
\end{table*}

\subsection{The Reflective Reception Network}
\label{subsec:reflection}

\subsubsection{\bf Projection Gate for Latent Embedding Learning} We first introduce the projection gate in our reflective reception network to map the user-item interaction data into the latent learning space. We denote the interaction vector of user $u_i$ and $t_j$ as $\textbf{R}_{i}\in\mathbb{R}^{N}$ ($[r_{i,1},...,r_{i,N}]$ and $\textbf{R}_{j}\in\mathbb{R}^{M}$ ($[r_{1,j},...,r_{M,j}]$, respectively. Without loss of generality, each column of $\textbf{R}$ is denoted as the interaction vector $\textbf{R}_{j}$ from the item perspective as the concrete example of vector dimension setting. Our framework is flexible to handle user's interaction input $\textbf{R}_{i}$ in a similar way. The projection gate operation is formally defined as: $\textbf{S}_{j} = \textbf{V}\cdot\textbf{R}_{j}$, where $\textbf{V}\in \mathbb{R}^{d_{p}\times N}$ is the learned transformation matrix and $d_p$ denotes the hidden state dimensionality ($d_p<<N$). $\textbf{S}_{j}$ is the latent representation of sparse user-item interaction data $\textbf{R}_{j}$.

\subsubsection{\bf Reflection Gate for Interaction Encoding}
To explore the different user-item relations and alleviate the variable sparsity bias, we further design a reflection gate to propagate the learned latent embeddings back to the original data space, which explicitly learns the unknown user-item pairwise preference scores in an effective manner. To achieve this goal, we first define a decay function $\phi(\cdot)$ which is analogous to the energy dissipation in the mirror reflection phenomenon to prevent the imputed preference scores learned from the reflection gate dominating the user-item interaction data. Given the interaction vector $\textbf{R}_{j}$, we formally present the decay function as: $\phi(\textbf{R}_{j}) = \frac{\alpha}{\|\textbf{R}_{j}\|_0}$
, where $\alpha$ is a hyperparameter controlling the decay strength and $\|\textbf{R}_{j}\|_0$ represents the number of observed interactions. It is worth mentioning that the decay function normalizes the energy of the reflected inputs using $\ell_0$ norm, as the sparsity degree varies by users/items causing the energy gaps in collaborative filtering architecture. The numerator can be substituted with other functions (\eg~$\log{(\|\textbf{R}_{j}\|_0+1)}$ and $\sqrt{\|\textbf{R}_{j}\|_0}$), to adjust how the reflection energy is normalized. In experiments, we will compare the performance of using the above decay functions. Based on the above definitions, the reflection gate operation can be formally given as:
\begin{align}
\label{equation:inference}
\hat{\textbf{R}}_{j} 
=\phi(\textbf{R}_{j}) \cdot \textbf{V}^\top \cdot \textbf{S}_{j}
=\phi(\textbf{R}_{j}) \cdot \textbf{V}^\top \cdot \textbf{V}\cdot \textbf{R}_{j}
\end{align}
\noindent where the projection weight matrix is denoted by $\textbf{V}^\top$, and $\hat{\textbf{R}}_{j}$ contains inferred quantitative preferences ($\hat{r}_{i,j}$ given $r_{i,j}=0$). Then we combine the inferred scores with the observed interactions:
\begin{align}
\label{equation:maskedAdd}
\tilde{\textbf{R}}_{j}
=h_1(\textbf{R}_{j};\{\textbf{V}\})
=\textbf{R}_{j}+o_{\textbf{R}_{j}}\circ \hat{\textbf{R}}_{j}
\end{align}
\noindent where $o_{\textbf{R}_{j}}$ is the zero indicator vector in which the element is set as 0 given the observed user-item interaction. $\circ$ denotes element-wise product operation.
Note that the higher the learned preference score $\hat{r}_{i,j}$ is, the more likely that user $u_i$ is interested in item $t_j$. The value of $\hat{r}_{i,j}$ is inferred with a smaller scale. The self-augmented interaction matrix $\tilde{\textbf{R}}$ not only preserves the original observed user-item interactions, but also encodes the relations between users and their non-interacted items. With the cooperation of reflection mechanism and decay factor, \model\ alleviates the data imbalance issue, \ie, dense interaction vectors often have higher energy than the sparse ones, by generating the reflected vectors with the uniform energy distribution regardless of the variable sparsity degrees.

\subsubsection{\bf Tied-Weight Scheme for Training Efficiency} In the reflective reception network, we integrate the projection and reflection gate with the tied-weight mechanism~\cite{kamyshanska2013autoencoder,luo2017convolutional}, \ie, the transformation operations are designed as transpose matrices. With the dedicated parameter sharing schema, the incorporated parameter learning constrain could effectively address the variable sparsity issue~\cite{zheng2016neural}. Such a learning scheme could improve the efficiency of \model\ by reducing its training cost and serve as an effective way of regularization to alleviate overfitting. In Section~\ref{sec:eval}, we perform ablation study by replacing $\textbf{V}^\top$ with another learnable transformation matrix $\textbf{U}\in\mathbb{R}^{N\times d_p}$, to show the effectiveness of our designed tied-weight scheme in \model\ framework.

\subsection{Collaborative Dependency Auto-Encoding}
To capture the inter-dependencies among different types of interactive relations (\ie, observed positive user-item interactions and learned unknown user-item interactions), we augment our \model\ framework with an information fusion autoencoder architecture. The developed autoencoder model endows the \model\ with the ability of fusing collaborative signals by learning the reconstruction of the recalibrated relation-aware user-item interactions $\tilde{\textbf{R}}$. In the process of our collaborative signal fusion, we first design an encoder to map the interaction matrix $\tilde{\textbf{R}}$ into another low-dimensional latent space which preserves the most informative signals across different relations. Following the encoder module, the decoder component aims to reconstruct $\tilde{\textbf{R}}$ from the encoded representation to be as close to the original input. The dimension of the output layer in our autoencoder network is the same as that of the input layer. The interaction relation fusion network is trained by minimizing the reconstruction loss between the original input and the corresponding reconstructed interactions, and jointly optimized with the reflective reception network.

We configure our collaborative dependency auto-encoding framework with a three-layer autoencoder framework, in which the predictive function is defined as follows:
\begin{align}
\label{equation:aepred}
h_2(\tilde{\textbf{R}}_{j};\{\textbf{W},\textbf{V},\textbf{c},\textbf{b}\})=\textbf{W}\cdot g(\textbf{V}\cdot\tilde{\textbf{R}}_{j} + \textbf{c}) + \textbf{b}
\end{align}
\noindent where $\textbf{W}\in\mathbb{R}^{N\times d_p}$, $\textbf{V}\in\mathbb{R}^{d_p\times N}$ are the transformation matrices, $\textbf{c}\in\mathbb{R}^{d_p}$, $\textbf{b}\in\mathbb{R}^N$ are the bias vectors, and $g(\cdot)$ is the sigmoid function such that $g(\textbf{x})(i)=g(\textbf{x}(i)), g(x)=\frac{1}{1+e^{-x}}$. Note that for the purpose of reducing models' parameters and achieving model efficiency, we conduct the parameter sharing mechanism by using the same transformation $\textbf{V}$ as the projection matrix in the hidden layer of the auto-encoder network.

\noindent \textbf{Parametric Regularized Reflection Process}. Furthermore, to correct the data imbalance bias with the flexibility in handling user's interaction vectors with various sparsity degrees, we further augment our \model\ by introducing a new parametric matrix $\textbf{T}\in\mathbb{R}^{d_p\times d_p}$ and update the predictive function of \model\ as:
\begin{align}
\label{equation:pred}
H(\textbf{R}_{j};\Theta)=\textbf{W}\cdot g(\textbf{V} \cdot \textbf{R}_{j}+\phi(\textbf{R}_{j})\cdot\textbf{T} \cdot \textbf{V} \cdot\ \textbf{R}_{j}+\textbf{c})+\textbf{b}
\end{align}
where $\textbf{T}$ replaces $\textbf{V}$ from the collaborative dependency auto-encoder and $\textbf{V}^\top$ from the reflective reception network. Then the following regularization term is incorporated in the loss function as follows:
\begin{align}
\label{equation:refreg}
\mathcal{L}_\text{reg}=\lambda_3\cdot\|\textbf{T}-\textbf{V} \cdot \textbf{V}\tp\|_\text{F}^2
\end{align}
The above regularization term guides the learning process of \model\ by simulating the reflection process in encoding the implicit user-item relations. Moreover, this design has another advantage in further optimizing the model training efficiency, \ie, $\textbf{T}$ is of the size $d_p\times d_p$ which is smaller than $d_p\times N$ and $N\times N$. Due to $d_p<<N$, it reduces two times of large-matrix multiplications to one-time small-matrix multiplication, which is more computationally efficient.

\subsection{Implicit Relation Learning with Reflective Autoencoder Framework}
\label{sec:relation}
With the integration of the reflective reception network and collaborative dependency autoencoder module, \model\ draws on the latent factor modeling, by transferring the positive user-item interactive signals to encode the user's implicit preferences on all other unobserved items. By incorporating the decay factor into reflective auto-encoder architecture, the imputed user's preference values has much smaller magnitude as compared to the original user-item interactions (\eg, $\hat{r}_{i,j}$ might be around 0.1 in a five-star rating system). Such an energy decay mechanism could prevent the dominating phenomenon caused by the imputed values, which misleads the prediction.

Different from the stacked autoencoder models~\cite{jiang2018stacked,wang2015relational}, our \model\ is a hierarchical framework with the designed reflective reception mechanism, which enables the parameter sharing and greatly reduces the model size. This model refining may decrease the difficulty of training and get rid of the greedy pre-train strategy, which has been applied in stacked autoencoder network~\cite{sedhain2015autorec}, to achieve the model efficiency. In addition, to further improve the model training performance and address the vanishing gradient problem, we configure \model\ with the residual connection strategy from the input to the output layer of \model.

\subsection{Model Optimization}

\begin{algorithm}[t]
    \footnotesize
	\caption{Learning Process of \model\ Framework}
	\label{alg:learn_alg}
	\LinesNumbered
	\KwIn{user-item interaction matrix $\textbf{R}\in\mathbb{R}^{N\times M}$, maximum epoch number $E$, regularization weights $\lambda_1, \lambda_2, \lambda_3$, learning rate $\eta$}
	\KwOut{trained parameters in $\Theta$}
	Initialize all parameters in $\Theta$\\
    \For{$e=1$ to $E$}{
        Draw a mini-batch $\mathcal{I}$ from all items $\{1,2,...,M\}$\\
        $\text{Loss} = \lambda_1\cdot\|\textbf{V}\|_\text{F}^2+\lambda_2\cdot\|\textbf{W}\|_\text{F}^2+\lambda_3\cdot\|\textbf{T}-\textbf{V}\cdot\textbf{V}^\top\|_\text{F}^2$\\
        \For{each $t_j\in\mathcal{I}$}{
            Compute $H(\textbf{R}_j;\Theta)$ according to Eq~\ref{equation:pred}\\
            $Loss += \|(1-o_{\textbf{R}_{j}})\circ(H(\textbf{R}_j;\Theta)-\textbf{R}_{j})\|_2^2$
        }
        \For{$\textbf{V},\textbf{W},\textbf{T}\in\Theta$}{
            $\textbf{V}=\textbf{V}-\eta\cdot\partial\text{Loss}/\partial\textbf{V}$\\
            $\textbf{W}=\textbf{W}-\eta\cdot\partial\text{Loss}/\partial\textbf{W}$\\
            $\textbf{T}=\textbf{T}-\eta\cdot\partial\text{Loss}/\partial\textbf{T}$
        }
    }
    return all parameters $\Theta$
\label{alg:optimization}
\end{algorithm}

Alg 1 presents the learning process of \model. Given the predictive function defined in Eq~\ref{equation:pred},
we perform the jointly training for both the reflective reception network and collaborative dependency autoencoder module with the following loss function:
\begin{align}
\label{equation:objfunc}
\min\limits_\Theta
\sum_{\textbf{R}_j\in \mathbb{T}} \|(\textbf{1}-o_{\textbf{R}_j})\circ (H(\textbf{R}_j;\Theta) - \textbf{R}_j)\|_2^2 + \lambda_1\cdot\|\textbf{V}\|_\text{F}^2 + \lambda_2\cdot \|\textbf{W}\|_\text{F}^2 + \lambda_3\cdot \|\textbf{T} - \textbf{V} \cdot \textbf{V}^\top\|_\text{F}^2\
\end{align}
where $\mathbb{T}$ denotes the training set, and $\textbf{1}=[1,\cdots,1]^\top\in\mathbb{R}^N$. The first component of loss function ($\|(1-o_{\textbf{R}})\circ (H(\textbf{R};\Theta) - \textbf{R})\|_2^2$) is the reconstruction term which is based on the observed user-item interactions. Hence, there is no restriction on the originally unobserved ratings, even they can be filled with the implicit user-item preference scores learned by the reflective reception network as defined in Eq~\ref{equation:inference}. The second component contains two weight decay regularization terms on $\textbf{V},\textbf{W}$ and the reflective loss defined by Eq~\ref{equation:refreg}. Scalars $\lambda_1,\lambda_2,\lambda_3$ are defined to balance the three terms. We apply the Adam optimizer to infer the model parameters.

\subsection{Theoretical Analysis}

Our reflection receptor simulates the energy propagation. The benefit of our reflection mechanism can be interpreted from two perspectives: (1) we compare the cosine similarity between the observed ratings and the imputed values in the latent space, and find that by adopting the transpose matrix $\textbf{V}^\top$ for the reflection process, the similarity is guaranteed to be positive. In other words, we prove that the similarity between the original data and the imputed data given by our approach, has a lower bound. Second, by comparing the transpose $\textbf{V}^\top$ with an independent variable $\textbf{U}$, we find that the equivalent solutions of our approach are the set of orthogonal matrices, and that of the trivial method are the invertible matrices. This makes the solutions to our method much less ambiguous and more stable.

One reasonable criterion on the implicit relation encoding of \model\ is that the imputed values $h_1(\textbf{R}_{j};\Theta_1)$ should be similar and consistent with the originally observed ratings $\textbf{R}_{j}$~\cite{xue2017deep}. Particularly, the consistency between $h_1(\textbf{R}_{j};\Theta_1)$ and $\textbf{R}_{j}$ can be measured by the similarity between their projected latent representations, \ie~$\textbf{V} \cdot h_1(\textbf{R}_{j};\Theta_1)$ and $\textbf{V} \cdot \textbf{R}_{j}$. To justify the rationality of reflection mechanism in \model, we first replace the transformation $\textbf{V}^\top$ with its counterpart $\textbf{U}$. Then we formally estimate the correlation between $\textbf{V} \cdot h_1(\textbf{R}_{j};\Theta_1)$ and $\textbf{V} \cdot \textbf{R}_{j}$ based on cosine similarity:
\begin{equation}
\label{equation:derivation}
\begin{split}
&\quad\,\cos(\textbf{V}\cdot \textbf{R}_{j}, \textbf{V} \cdot h_1(\textbf{R}_{j};\Theta_1))=\cos(\textbf{V} \cdot \textbf{R}_{j}, \phi(\textbf{R}_{j})\cdot \textbf{V} \cdot \textbf{U} \cdot \textbf{V} \cdot \textbf{R}_{j})
=\cos(\textbf{V} \cdot \textbf{R}_{j}, \textbf{V} \cdot \textbf{U} \cdot \textbf{V} \cdot \textbf{R}_{j})\\
&=\frac{(\textbf{R}_{j}\tp \cdot \textbf{V}\tp)(\textbf{V} \cdot \textbf{U} \cdot \textbf{V} \cdot \textbf{R}_{j})}{\|\textbf{V} \cdot \textbf{R}_{j}\|_2 \cdot \|\textbf{V} \cdot \textbf{U} \cdot \textbf{V} \cdot \textbf{R}_{j}\|_2}
=\frac{(\textbf{V}\tp \cdot \textbf{V} \cdot \textbf{R}_{j})\tp(\textbf{U} \cdot \textbf{V} \cdot \textbf{R}_{j})} {\|\textbf{V} \cdot \textbf{R}_{j}\|_2 \cdot \|\textbf{V} \cdot \textbf{U} \cdot \textbf{V} \cdot \textbf{R}_{j}\|_2}\\
&=\frac{\|\textbf{V}\tp \cdot \textbf{V} \cdot \textbf{R}_{j}\|_2 \cdot \|\textbf{U} \cdot \textbf{V} \cdot \textbf{R}_{j}\|_2} {\|\textbf{V} \cdot \textbf{R}_{j}\|_2 \cdot \|\textbf{V} \cdot \textbf{U} \cdot \textbf{V} \cdot \textbf{R}_{j}\|_2} \cdot \cos(\textbf{V}\tp \cdot \textbf{V} \cdot \textbf{R}_{j}, \textbf{U} \cdot \textbf{V} \cdot \textbf{R}_{j}).
\end{split}
\end{equation}
Since $\frac{\|\textbf{V}\tp \cdot \textbf{V} \cdot \textbf{R}_{j}\|_2 \cdot \|\textbf{U} \cdot \textbf{V} \cdot \textbf{R}_{j}\|_2} {\|\textbf{V} \cdot \textbf{R}_{j}\|_2 \cdot \|\textbf{V} \cdot \textbf{U} \cdot \textbf{V} \cdot \textbf{R}_{j}\|_2}$ is non-negative, the sign in Eq~\eqref{equation:derivation} is dependent on the sign of $\cos(\textbf{V}\tp \cdot \textbf{V}\cdot \textbf{R}_{j}, \textbf{U}\cdot \textbf{V} \cdot \textbf{R}_{j})$. For a free-form $\textbf{U}$, it is likely that $\cos(\textbf{V} \cdot \tp\textbf{V} \cdot \textbf{R}_{j},\textbf{U} \cdot \textbf{V} \cdot \textbf{R}_{j})$ is negative which makes $\textbf{V} \cdot \textbf{R}_{j}$ and $\textbf{V} \cdot h_1(\textbf{R}_{j};\Theta_1)$ opposite to each other. In contrast, by setting the transformation $\textbf{U}=\textbf{V}^\top$ with the tied-weight scheme, we have
\begin{equation}
\cos(\textbf{V}\tp \cdot \textbf{V} \cdot \textbf{R}_{j},\textbf{U} \cdot\textbf{V} \cdot \textbf{R}_{j})=\cos(\textbf{V}\tp \cdot \textbf{V} \cdot \textbf{R}_{j},\textbf{V}\tp \cdot \textbf{V} \cdot \textbf{R}_{j})=1
\end{equation}
Then, the cosine similarity in Eq~\eqref{equation:derivation} is guaranteed to be positive and thus the reflected vectors do not contain the components which are opposite to the original user-item interaction matrix.

The benefit of using $\textbf{V}^\top$ instead of $\textbf{U}$ can also be interpreted from the perspective of solution's ambiguity. Suppose $(\textbf{U}^*,\textbf{V}^*)$ is one solution for \model, $(\textbf{U}^*\textbf{B}^{-1},\textbf{B} \cdot \textbf{V}^*)$ is an equivalent solution based on the following formal operations:
\begin{equation}\label{eq:amb_sol}
\begin{split}
\quad\,h_1(\textbf{R}_{j};\{\textbf{U}^*,\textbf{V}^*\})=\phi(\textbf{R}_{j})\cdot \textbf{U}^*\textbf{V}^*\textbf{R}_{j}
=\phi(\textbf{R}_{j})\cdot \textbf{U}^*\textbf{B}^{-1}\cdot\textbf{B}\textbf{V}^*\textbf{R}_{j} =h_1(\textbf{R}_{j};\{\textbf{U}^*\textbf{B}^{-1},\textbf{B}\textbf{V}^*\})
\end{split}
\end{equation}
Thus, we have
\begin{equation}\label{eq:amb_sol_oth}
\mathbb{S}_1=\{(\textbf{U}^*\textbf{B}\inv, \textbf{B}\textbf{V}^*):\textbf{B}^{-1}\,\text{exists}\}; \mathbb{S}_2=\{({\textbf{V}^*}\tp \textbf{B}\tp, \textbf{B} \textbf{V}^*):\textbf{B}\textbf{B}^\top=\textbf{B}^\top\textbf{B}=\textbf{I}\}
\end{equation}
\noindent $\mathbb{S}_1$ is a set of equivalent solutions for \model\ model. We consider $\textbf{U}=\textbf{V}^\top$ and suppose $({\textbf{V}^*}\tp,\textbf{V}^*)$ is one solution. Then the corresponding set of equivalent solutions $\mathbb{S}_2$ is also given above.


Note that $\|\mathbb{S}_1\|$ is much larger than $\|\mathbb{S}_2\|$, as orthogonal matrices are only a small subset of invertible matrices. In other words, replacing $\textbf{V}^\top$ by $\textbf{U}$ in \model\ would increase the ambiguity of solution of the model. In addition, an invertible matrix may have poor condition so that some solutions in $\mathbb{S}_1$ affect the stability of the module, while any orthogonal matrices are well-posed so that any solution from $\mathbb{S}_2$ is equivalently stable for the network.

\subsection{Complexity Analysis of the \model\ Framework}
As discussed in Section~\ref{sec:relation}, although the proposed \model\ model seems to improve the performance by introducing more transformation terms, most of the parameters in the reflection mechanism are shared by the existing parameters of autoencoder. Specifically, the number of parameters of \model\ with tied weights is $2N\times d_p$ which is the same as that of a three-layer auto-encoder, and the model size of \model\ with implicit reflection (\ie, utilizing $\textbf{T}$ with $\mathcal{L}_\text{reg}$) is $(2N\times d_p + d_p\times d_p)$ which costs moderate extra space considering $d_p<<N$. As for the computational cost, $2N\times d_p$ and $d_p\times d_p$ of extra computations are required, compared to the cost of $2N\times d_P$ with a three-layer autoencoder. In conclude, the reflection mechanism does not change the order of both the time and the space complexity when it is built into an autoencoder framework. Specifically, the time and the space complexity are all $O(N\times d_p)$, which is competitive as compared to many neural network-based approaches. We further investigate the computational cost of our developed \model\ in Section~\ref{sec:model_efficiency}, to show the model efficiency of \model\ as compared to state-of-the-art collaborative filtering techniques.

%% file: eval.tex
\section{Evaluation}
\label{sec:eval}

We perform extensive experiments to evaluate the performance of \emph{\model} on several real-world datasets, and make comparison with different types of state-of-the-art techniques. Specifically, we aim to answer the following research questions:

\begin{itemize}[leftmargin=*]

\item \textbf{RQ1}: How does \model\ perform as compared with state-of-the-art recommendation methods on different datasets?\\\vspace{-0.1in}

\item \textbf{RQ2}: How is the performance of \model\ \wrt\ different reflection mechanism in the reflective reception network?\\\vspace{-0.1in}

\item \textbf{RQ3}: How does the designed reflection reception network contribute to the recommendation performance of \model?\\\vspace{-0.1in}

\item \textbf{RQ4}: How is the impact of different decay functions for the model recommendation performance?\\\vspace{-0.1in}

\item \textbf{RQ5}: How is \emph{\model}'s forecasting performance \wrt\ different data sparsity degrees?\\\vspace{-0.1in}

\item \textbf{RQ6}: How is the model performance \wrt\ user-based and item-based collaborative filtering architecture?\\\vspace{-0.1in}

\item \textbf{RQ7}: How is the model efficiency of our \model\ framework in handling different scales of data?\\\vspace{-0.1in}

\item \textbf{RQ8}: What is the hyperparameter influences for the performance of our \model\ framework?

\end{itemize}


\begin{table*}
	\centering
	\scriptsize
	\vspace{-0.1in}
	\caption{Statistics of Experimented Datasets.}
	\vspace{-0.15in}
	\label{tab:dataset}
	\setlength{\tabcolsep}{1mm}
	\begin{tabular}{cccccccccccc}
		\toprule
		Dataset&Density&
		&User \#
		&\tabincell{c}{Range of\\User Rating \#}
		&\tabincell{c}{AVG User\\Rating \#}
		&\tabincell{c}{{\it s.t.d.} of User\\Rating \#}&
		&Item \#
		&\tabincell{c}{Range of\\Item Rating \#}
		&\tabincell{c}{AVG Item\\Rating \#}
		&\tabincell{c}{{\it s.t.d.} of Item\\Rating \#}\\
		\hline
		ML-1M&0.05&&6040&20$\sim$2314&165.60&192.73&
		&3706&1$\sim$3428&269.89&383.99\\
		
		ML-10M&0.01&&69878&20$\sim$7359&143.11&216.71&
		&10677&1$\sim$34864&936.60&2487.21\\
		
		Netflix&0.01&&480189&1$\sim$17653&209.25&302.34&
		&17770&3$\sim$232944&5654.50&16909.20\\
		
		Foursquare&0.005&&24748&5$\sim$244&40.22&21.03&
		&7763&5$\sim$7741&128.22&359.84\\
		
		Yelp & 0.007 && 11894 & 20$\sim$1212 & 58.56 & 50.19 && 8560 & 10$\sim$782 & 81.37 & 52.23\\
		\hline
	\end{tabular}
\end{table*}

\subsection{Experiment Settings}
\subsubsection{\bf Data Description}
We evaluate the proposed model on four benchmark recommendation datasets: MovieLens-1M (ML-1M), MovieLens-10M (ML-10M), Netflix prize data and Foursquare data, corresponding to two different recommendation tasks: \emph{rating prediction} and \emph{item recommendation}. Table~\ref{tab:dataset} summarizes the data statistics. 

\noindent \textbf{Online Rating Data}. These three rating datasets (\ie, ML-1M\footnote{https://grouplens.org/datasets/movielens/1m/}, ML-10M\footnote{https://grouplens.org/datasets/movielens/10m/} and Netflix\footnote{https://www.kaggle.com/netflix-inc/netflix-prize-data}), which vary in data sparsity degrees and user/item scales, have been widely used for evaluating various recommender systems. In particular, the user-item interaction data is formatted as (user id, movie id, rating score). Each rating score of ML-1M and Netflix data ranges from 1 to 5 stars with one star as increments, and the ML-10M data comes from the five star system using half-star as the increment. The preference matrices of MovieLens-1M (ML-1M), MovieLens-10M (ML-10M) and Netflix prize data, is on the scale of $10^3\times 10^3$, $10^4\times 10^4$ and $10^5\times10^4$, respectively.

\noindent \textbf{Foursquare Check-in Data}\footnote{https://sites.google.com/site/yangdingqi/home/foursquare-dataset?authuser=0}. It collects the mobility traces of users by recording their check-ins over different venues. Following the similar settings in~\cite{yang2019revisiting}, we remove users and items whose check-in interactions are less than 20.

\noindent \textbf{Yelp Data}\footnote{https://www.yelp.com/dataset/download}. This dataset has been widely used for evaluating the performance of developed recommender systems. It is collected from the Yelp platform for business venue recommendation.

\subsubsection{\bf Evaluation Protocols}

In our experiments, we divide each experimented dataset into training and test set with the ratio of 9:1. Specifically, 5\% of the training set is used as the validation set for hyperparameter tuning and the validation set determines the parameters which generate the best performance. All experiments were conducted for five runs with dataset re-division. The averaged performance is reported in the evaluation results.

To evaluate the model performance, we adopt two types of evaluation metrics for different recommendation tasks.

\noindent \textbf{Rating Prediction}. We employ the Root Mean Squared Error (RMSE) as the rating prediction evaluation metric which is formally defined as follows:
\begin{equation}
RMSE=\sqrt{\frac{\sum_{r\in \mathbb{S}}
		\|(1-o_{r})\circ(r-\hat{r})\|_2^2}
	{\sum_{r\in \mathbb{S}}\|r\|_0}}.
\end{equation}
\noindent where $r$ and $\hat{r}$ represents the actual and predicted rating score, respectively. The validation set is denoted as $\mathbb{S}$. During the test phase, the prediction ability of the model, instead of the reconstruction ability, is evaluated by using the vectors from the training set as inputs. The lower RMSE value indicates better performance. Through all the experiments, the standard deviation on each five RMSE scores is within $0.002$ by our method which is rather small. Thus, we omit the standard deviation in the reported results.

\noindent \textbf{Item Recommendation}. We use two popular metrics for ranking-based performance measurement \emph{Precision@$K$} and \emph{Normalized Discounted Cumulative Gain (NDCG@$K$)}~\cite{fan2019deep}, to evaluate the accuracy of item recommendations. Multiple $K$ values are applied for evaluation. Note that a higher Precision and NDCG values reflect better performance.

\subsubsection{\bf Methods for Comparison}
To comprehensively evaluate our \model, we compare it with different baselines from five research lines for comparison: 1) conventional matrix factorization recommender systems; 2) neural collaborative filtering models for recommendation; 3) learning to recommend with graph neural networks; 4) neural auto-regressive recommendation models; 5) autoencoder collaborative filtering recommendation techniques.

\noindent \textbf{Conventional Matrix Factorization Recommender Systems}: We first consider several representative conventional baselines which are built based on the matrix factorization architecture.
\begin{itemize}[leftmargin=*]
\item \textbf{PMF}~\cite{mnih2008probabilistic}: It is a representative probabilistic recommendation algorithm that projects the user preference matrix into lower rank sub-matrices for users and items with the Gaussian noises.
\item \textbf{SVD}++~\cite{koren2008factorization}: This model combines the matrix factorization architecture with neighborhood-based methods, by integrating the latent factors and neighborhood signals into the framework.
\item \textbf{BiasMF}~\cite{koren2009matrix}: It enhances the matrix factorization \wrt\ the model flexibility by incorporating user and item biases.
\item \textbf{LORM}~\cite{lee2013local}: It relaxes the low-rank assumption on the interaction matrix with the local matrix factorization.
\end{itemize}

\noindent \textbf{Neural Collaborative Filtering Models for Recommendation}: We further compare \emph{\model} with neural network-enhanced collaborative filtering methods through multilayer perceptron.
\begin{itemize}[leftmargin=*]
\item \textbf{DMF}~\cite{xue2017deep}: It is a neural matrix factorization model which learns latent vectors for both users and items, with the consideration of both explicit ratings and implicit feedback.
\item \textbf{NCF}~\cite{he2017neural}: we consider three implementation versions of NCF method with different concatenation schemes: \ie, NCF-G: using the element-wise production; NCF-M: leveraging the Multilayer perceptron for embedding projection; NCF-N: applying the element-wise-product branch.


\end{itemize}

\noindent \textbf{Neural Auto-regressive Recommendation Models}: \emph{\model} is also compared against auto-regressive recommendation models with neural network architectures.
\begin{itemize}[leftmargin=*]
\item \textbf{RBM}~\cite{salakhutdinov2007restricted}: RBM is a representative pioneering neural network-based model which apply the classic restricted boltzmann machines to the collaborative filtering architecture.
\item \textbf{NADE}~\cite{zheng2016neural}: It is a neural autoregressive network for the collaborative filtering with the the parameter sharing.
\item \textbf{CF-UIcA}~\cite{du2018collaborative}: It simultaneously models users' and items' rating vectors based on the neural autoregressive framework.
\end{itemize}

\noindent \textbf{Autoencoder Collaborative Filtering Architecture}: One key research line of collaborative filtering models lies in the utilization of autoencoder for mapping user-item interactions into latent representations. \emph{\model} also competes with the autoencoder CF paradigm.
\begin{itemize}[leftmargin=*]
\item \textbf{ACF}~\cite{ouyang2014autoencoder}: The method is an autoencoder-based collaborative filtering model which regards each rating score as a unique class in the recommendation scenario.
\item \textbf{AutoRec}~\cite{sedhain2015autorec}: The model learns compressed user/item representations by designing a three-layer autoencoder architecture in which rating scores treated as continuous values.
\item \textbf{CDAE}~\cite{strub2016hybrid}: It debiases the auto-encoder with an adaptive loss function to learn robust user/item representations.
\end{itemize}

\noindent \textbf{Recommendation with Graph Neural Networks}: Another important relevant research line is to explore the user-item interactive relation structure in recommender systems. Three recent developed graph neural network models are considered in the performance comparison.
\begin{itemize}[leftmargin=*]
\item \textbf{GC-MC}~\cite{berg2017graph}: It is a graph-based autoencoder architecture that leverages the graph convolutions to jointly handle the user-item interactions and auxiliary information.

\item \textbf{ST-GCN}~\cite{zhang2019star}: It leverages the graph convolutional network as the aggregator in the encoder-decoder framework to learn the node embeddings for recommendation.

\item \textbf{NGCF}~\cite{wang2019neural}: This approach learns the vector representations of users and items, by utilizing the graph neural network to perform the propagation for collaborative signal encoding.
\end{itemize}

\noindent \textbf{Imputation-based Recommender Systems}: We further compare \emph{\model} with the line of collaborative filtering which are on the basis of data imputation techniques:
\begin{itemize}[leftmargin=*]

\item \textbf{S-Imp}~\cite{mazumder2010spectral}: this method imputes the missing elements with the computation of low-rank Singular Value Decomposition (SVD). In S-Imp, grid value-based regularization is applied in the framework.

\item \textbf{I-NMF}~\cite{wan2019sparse}: It is proposed to alleviate the data sparsity problem with the denoising autoencoder. The imputation-based neural matrix factorization model adopts the multiple-layer perceptron to optimize the collaborative filtering architecture.

\end{itemize}

\subsubsection{\bf Parameter Settings}
We implement all models with TensorFlow and utilize Adam as the model optimizer. The dimension of the hidden layer is set to $500$ which follows the settings in~\cite{sedhain2015autorec, zheng2016neural}. When $\textbf{U}$ is used to replace $\textbf{V}\tp$ in \model, $\lambda_4\|\textbf{U}\|_\text{F}^2$ is added to the loss function~\eqref{equation:objfunc}. In the hyperparameter tuning, the regularization weights $\lambda_1$, $\lambda_2$, $\lambda_3$, $\lambda_4$ are selected from $\{$0.5, 0.1, 0.05, 0.01, 0.005, 0.001, 0.0005$\}$, and the hyperparameter $\alpha$ in decay function is selected from \{200, 20, 2, 0.2, 0.02\}. To be consistent with the implementation of graph neural recommendation models (ST-GCN and NGCF), the depth of embedding propagation layers for those baselines is searched from the value range of $\{1,2,3,4\}$.



\begin{table}
\small
\vspace{-0.05in}
\caption{Performance comparison of rating prediction.}
\vspace{-0.15in}
\label{tab:res_ml1m}
\centering
\setlength{\tabcolsep}{1.5mm}
\begin{tabular}{clccc}
	\hline
	Category & Methods & ML-1M & ML-10M & Netflix\\
	\hline
	\multirow{4}{*}{MF-based Model}
	&PMF&0.883&0.791&0.902\\
	&LLORMA&0.865&0.822&0.874\\
	&SVD++&0.852&0.787&0.887\\
	&BiasMF&0.845&0.803&0.844\\
	\hdashline
	\specialrule{0em}{1pt}{1pt}
	\multirow{4}{*}{Neural-CF}
	&DMF&0.845&0.792&0.839\\
	&NCF-MLP&0.866&0.799&0.825\\
	&NCF-GMF&0.848&0.788&0.828\\
	&NCF-NeuMF&0.840&0.780&0.818\\
	\hdashline
	\specialrule{0em}{1pt}{1pt}
	\multirow{3}{*}{GNN Method}
	&GC-MC&0.832&0.777&-\\
	&NGCF&0.830&0.778&-\\
	&ST-GCN&0.832&0.770&-\\
	\hdashline
	\specialrule{0em}{1pt}{1pt}
	\multirow{3}{*}{Neural Auto-regressive}
	&RBM&0.854&0.823&0.845\\
	&NADE&0.829&0.771&0.803\\
	&CF-UIcA&0.823&0.769&\textbf{0.799}\\
	\hdashline
	\specialrule{0em}{1pt}{1pt}
	\multirow{3}{*}{Auto-Encoder CF}
	&ACF&0.890&0.803&0.831\\
	&AutoRec&0.831&0.782&0.823\\
	&CDAE&0.833&0.778&0.819\\
	\hdashline
	\specialrule{0em}{1pt}{1pt}
	\multirow{2}{*}{Re-Imputation Method}
	& S-Imp & 0.838 & 0.783 & 0.831\\
	& I-NMF & 0.828	& 0.777 & 0.818\\
	\hline
	\specialrule{0em}{1pt}{1pt}
	Our Model &\emph{\emph{\model}}&\textbf{0.822}&\textbf{0.768}&0.814\\
	\hline
	\vspace{-0.2in}
\end{tabular}
\end{table}


\subsection{Performance Comparison (RQ1)}

\subsubsection{\bf Rating Prediction Task}
We first present overall evaluation results of our \emph{\model} in rating prediction task with detailed discussions as follows.

\noindent \textbf{Overall Performance of \emph{\model}}.
We evaluate the performance of all compared methods and report the results in Table~\ref{tab:res_ml1m}. From this table, we can observe that \emph{\model} outperforms different types of baselines on various datasets in most cases, which demonstrates the effectiveness of \emph{\model}. While two neural autoregressive methods (\ie, NADE and CF-UIcA) achieve slightly better performance on the Netflix dataset, we note that NADE is a more complicated model which involves more learnable parameters and is more time-consuming as compared to \model\ (the model scalability is investigated in later subsection). Due to the high computational and space cost of graph aggregation operation in GNN-based baselines (\ie, ST-GCN, GC-MC, NGCF), it is difficult to scale them on large-scale data (\ie, Netflix), \ie, ``-'' indicates the out of memory issue.

Additionally, another interesting observation is that the performance gap between our \model\ and other baselines becomes larger as the the data volume decreases, \ie, the Movielens rating data contains much less ratings compared to the Netflix dataset as shown in Table~\ref{tab:res_ml1m}. This observation justifies that \model\ is capable of effectively capturing latent relation structures between users and items with the reflection mechanism, in order to alleviate the data scarcity issue in the recommendation scenario.

\noindent \textbf{Comparison among Baselines}.
The relative improvement between neural models and conventional matrix factorization approaches, suggests that the neural architecture is able to capture non-linearity among user-item interactions. Among all the baselines, neural probabilistic approaches and autoencoder-based collaborative filtering outperform the conventional CF techniques as well as the neural CF models with the integration of MLP, suggesting that autoencoder and neural probabilistic models are more beneficial for modeling non-linear user-item interactions as compared to the simple MLP module. In the occasional cases that \model\ misses the best performance, it still generates very competitive results on the Netflix dataset when compared with two NADE-based models. However, we note that NADE-based methods are more complicated models which involve more learnable parameters and require larger computational cost during the training phase.

\begin{figure*}
    \centering
    \includegraphics[width=0.98\textwidth]{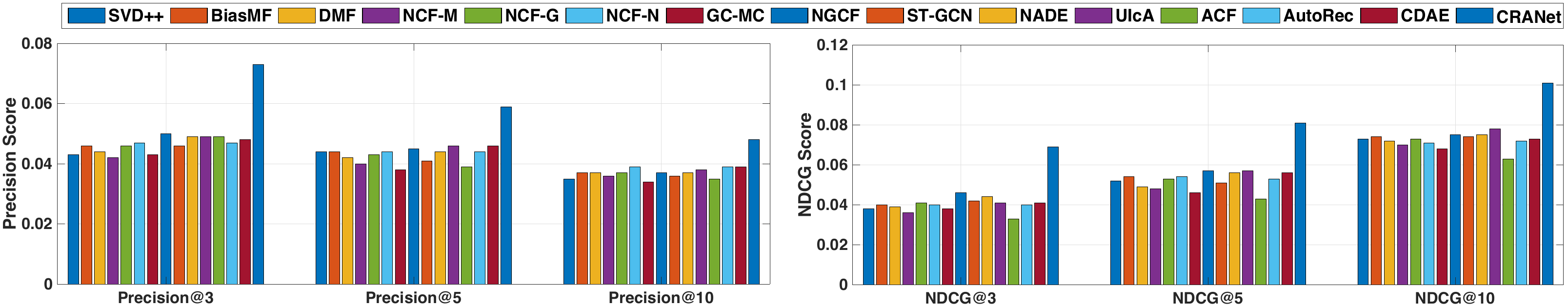}
    \vspace{-0.05in}
    \caption{Performance validation for location prediction in terms of \emph{Precision@$K$} and \emph{NDCG@$K$}.}
    \vspace{-0.05in}
    \label{fig:location_rec_fig}
\end{figure*}


\begin{table}
    \vspace{-0.05in}
	\caption{Performance comparison for item recommendation in terms of \textit{Precision@3} and \textit{NDCG@3}.}
	\vspace{-0.15in}
	\centering
    \scriptsize
	\setlength{\tabcolsep}{0.6mm}
    \begin{tabular}{|c|c|c|c|c|c|c|c|c|c|c|c|c|c|c|c|c|}
        \hline
        Model & Metrics & SVD++ & BiasMF & DMF & NCF & GC-MC & NGCF & ST-GCN & NADE & UIcA & ACF & AutoRec & CDAE & S-Imp & I-NMF & \emph{\model}\\
        \hline
        \multirow{2}{*}{ML-1M} & Prec@3 & 6.57\% & 6.58\% & 7.39\% & 8.62\% & 7.01\% & 9.18\% & 7.64\% & 13.81\% & 14.96\% & 7.92\% & 8.21\% & 8.65\% & 6.89\% & 8.44\% & \textbf{15.71\%}\\
        \cline{2-17}
        & NDCG@3 & 2.49\% & 2.49\% & 3.05\%& 3.51\% & 2.91\% & 3.84\% & 3.27\% & 5.51\% & 5.96\% & 3.01\% & 3.38\% & 3.51\% & 2.74\% & 3.44\% & \textbf{6.26\%}\\
        \hline
        
        \multirow{2}{*}{Fsq} & Prec@3 & 4.32\% & 4.57\% & 4.40\% & 4.63\% & 4.30\% & 4.98\% & 4.64\% & 4.93\% & 4.88\% & 4.02\% & 4.68\% & 4.81\% & 4.66\% & 4.75\% & \textbf{7.27\%}\\
        \cline{2-17}
        & NDCG@3 & 3.82\% & 4.03\% & 3.85\% & 4.08\% & 3.76\% & 4.60\% & 4.19\% & 4.44\% & 4.11\% & 3.26\% & 4.05\% & 4.15\% & 4.10\% & 4.10\% & \textbf{6.91\%}\\
        \hline
        
        \multirow{2}{*}{Yelp} & Prec@3 & 2.23\% & 2.42\% & 2.49\% & 2.50\% & 2.58\% & 2.73\% & 2.58\% & 2.94\% & 3.01\% & 2.72\% & 2.83\% & 2.79\% & 2.59\% & 2.93\% & \textbf{3.21\%}\\
        \cline{2-17}
        & NDCG@3 & 1.50\% & 1.58\% & 1.73\% & 1.84\% & 1.77\% & 1.83\% & 1.80\% & 2.01\% & 2.03\% & 1.81\% & 2.01\% & 1.97\% & 1.82\% & 2.04\% & \textbf{2.21\%}\\
        \hline
    \end{tabular}
    \label{tab:item_rec}
\end{table}

\subsubsection{\bf Item Recommendation Task}
We further perform experiments to evaluate all methods on three datasets (i.e., ML-1M, Foursquare and Yelp) for the item recommendation task. As shown in Table~\ref{tab:item_rec}, we can observe that we can observe that our proposed method \emph{\model} achieves the best performance as compared to various baselines, which demonstrates the effectiveness of our reflection-augmented autoencoder network in capturing the implicit users' preference from their unobserved interactions. Additionally, Figure~\ref{fig:location_rec_fig} shows the ranking performance of top-$K$ recommended locations on Foursquare dataset where $K\in \{3,5,10\}$. We can see that \emph{\model} achieves the best performance in both metrics Precision and NDCG in all cases, which points to the positive effect of modeling implicit users' preference over their non-interacted locations. Furthermore, while ST-GCN and NGCF utilize the high-order neighbor information to guide the representation learning of user-location interactions, they simply treat the unobserved interactions with zero imputation, which may lead to suboptimal performance. This further verifies the importance of capturing cross-modal collaborative signals in the embedding function of our reflective collaborative filtering framework.

\subsection{Impact of Reflection Mechanism Configuration in \emph{\model} (RQ2)}
To study the impact of different reflective reception network design in \emph{\model}, we perform experiments with different reflection mechanism configurations (an independent transformation operation $\textbf{U}$, transpose-based reflection gate $\textbf{V}\tp$ and regularization-based reflection gate $\textbf{T}$ with $\mathcal{L}_\text{reg}$). We also evaluate the performance by incorporating $\textbf{T}$ but without the regularization term defined in Eq.~\eqref{equation:refreg}. For fair comparison, when perform experiments on the same dataset, all models share the same hyperparameters $\lambda$ and $\alpha$. We also apply the regularization $\|\textbf{U}\|^2_{\text{F}}$ in the independent transformation.

\begin{figure*}
    \centering
    \includegraphics[width=1.0\textwidth]{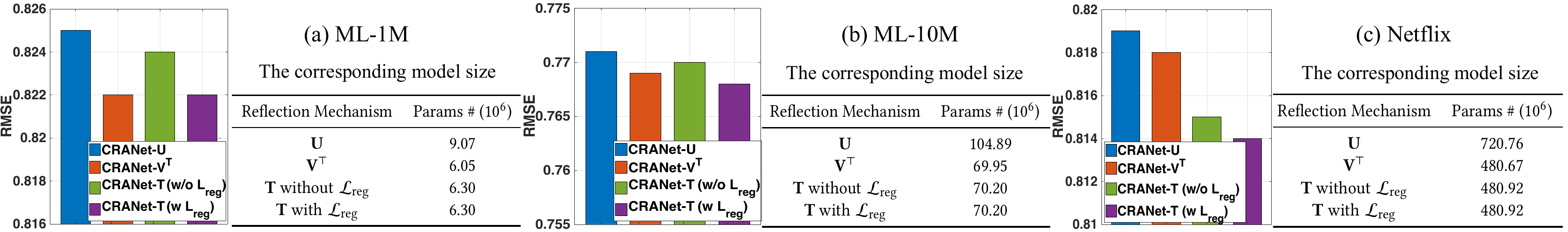}
    \caption{Recommendation performance of \emph{\model} with different reflection mechanism configurations.}
    \label{fig:RefMatTests}
\end{figure*}

Figure~\ref{fig:RefMatTests} shows the performance and corresponding model sizes. We observe that our regularization-based reflection mechanism $\textbf{T}$ with $\mathcal{L}_\text{reg}$ achieves the best performance when competing with other alternatives. In addition, the performance gap between $\textbf{T}$ with $\mathcal{L}_\text{reg}$ and $\textbf{V}\tp$ increases, as the average item density increases (see Table~\ref{tab:dataset}), which suggests that the transpose-based reflection gate $\textbf{V}\tp$ might over-regularize the model on large datasets and the incorporation of a lightweight regularization operation ($\textbf{T}$ with $\mathcal{L}_\text{reg}$) could alleviate this overfitting issue. Finally, it is worth mentioning that both $\textbf{V}\tp$ and $\textbf{T}$ have very low memory cost, given $d_p<<N$. Therefore, our \emph{\model} can be easily scaled to the largest Netflix dataset with efficient model training, while the independent transformation $\textbf{U}$ has high memory consumption.

\subsection{Model Ablation Study (RQ3)}
In this subsection, we perform model ablation study to investigate the effectiveness of our designed reflective reception network from the following two perspectives:
\begin{itemize}[leftmargin=*]
\item i) \emph{\model}-R: we remove the reflective reception network from the \emph{\model} and directly utilize a three-layer autoencoder network to learn latent representations of user-item interaction $\textbf{R}$.
\item ii) \emph{\model}-N: we replace the reflective reception network with a neighborhood based data imputation approach based on a similarity estimation scheme across users and items~\cite{desrosiers2011comprehensive}.
\end{itemize}
The experimental results are reported in Table~\ref{tab:reflection_reception_ablation}. With the help of our reflective reception network, our \emph{\model} achieves the best performance as compared to \emph{\model}-R and \emph{\model}-N.
The significant performance improvement suggests the effectiveness of our reflective reception module in capturing users' dynamic preferences over their non-interacted items, with an inner neural network structure to avoid the misleading of the user-item relation learning.

\subsection{Effect of Decay Function (RQ4)}
\begin{figure}
    \centering
    \subfigure[][Decay Func. Test on ML-1M]{
        \centering
        \includegraphics[width=0.32\textwidth]{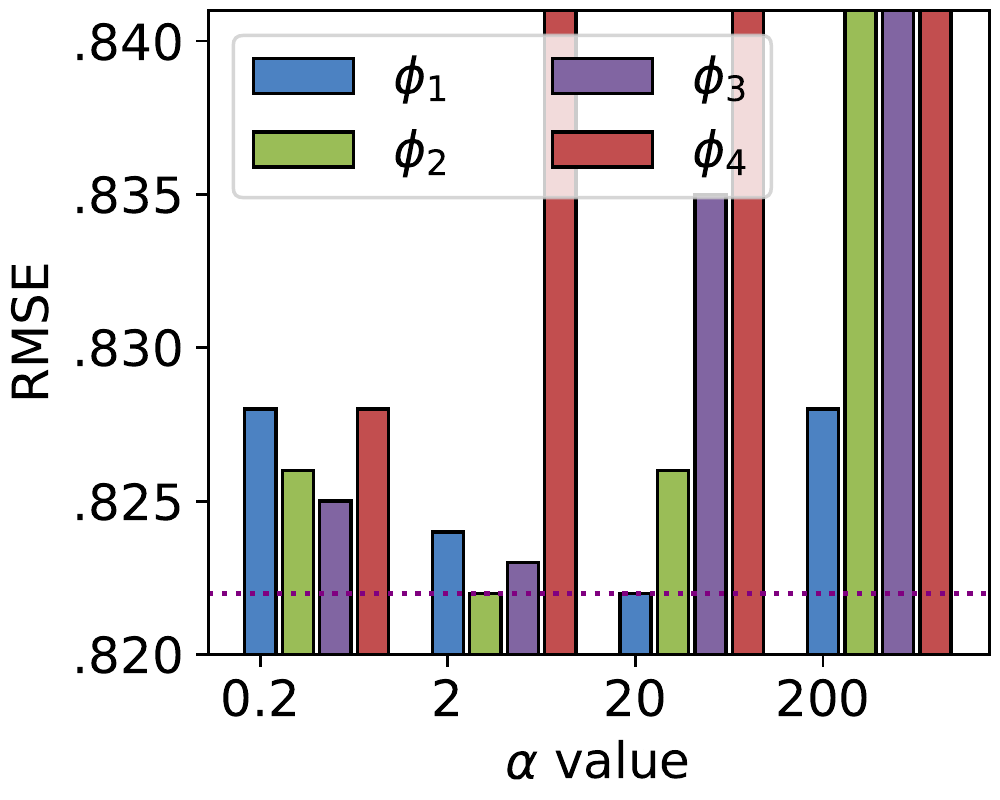}
        \label{fig:decay_ml1m}
        }
    \subfigure[][Decay Func. Test on ML-10M]{
        \centering
        \includegraphics[width=0.32\textwidth]{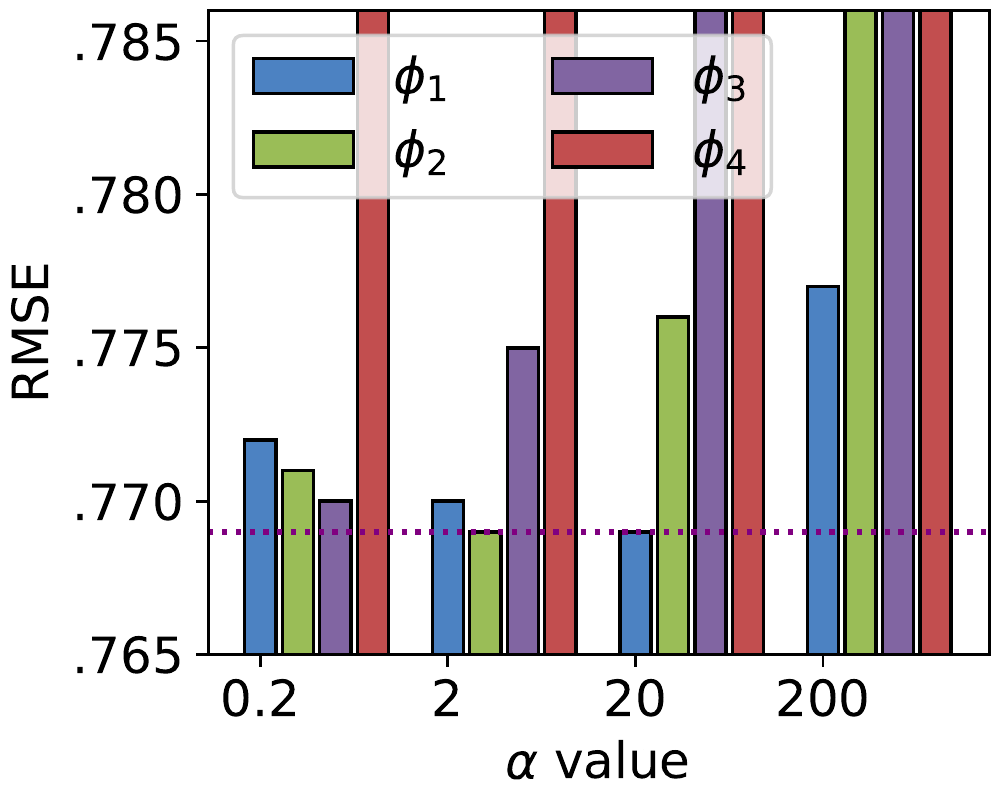}
        \label{fig:decay_ml10m}
        }
	\caption{Model Performance v.s. decay function settings on Movielens-1M and Movielens-10M Data}
    \label{fig:intTest}
    \vspace{-0.1in}
\end{figure}

We perform experiments to investigate the impact of decay function $\phi(\textbf{R}_{j})$ selection in the proposed reflective reception network of \model. In particular, recall four choices of the numerator in decay function mentioned in Sec~\ref{subsec:reflection}. Different choices result in the following different decay function candidates in formal presentations:
\begin{equation}
\label{equation:intensities}
\begin{split}
\phi_1(\textbf{R}_{j}) &= \frac{\alpha}{\|\textbf{R}_{j}\|_0}\quad
\phi_2(\textbf{R}_{j}) = \frac{\alpha\log{(\|\textbf{R}_{j}\|_0+1)}}{\|\textbf{R}_{j}\|_0}\quad \nonumber\\
\phi_3(\textbf{R}_{j}) &= \frac{\alpha\sqrt{\|\textbf{R}_{j}\|_0}}{\|\textbf{R}_{j}\|_0}\quad
\phi_4(\textbf{R}_{j}) = \frac{\alpha\|\textbf{R}_{j}\|_0}{\|\textbf{R}_{j}\|_0} = \alpha
\end{split}
\end{equation}
\noindent In the experiments, the tied weight-based reflection mechanism is employed. Figure~\ref{fig:decay_ml1m} and Figure~\ref{fig:decay_ml10m} shows the recommendation results with respect to the parameter $\alpha$, and Table~\ref{tab:best_perform_decay} summarizes the best performance of the four candidates with the corresponding $\alpha$ value. For all candidates except that using $\phi_4$, a proper $\alpha$ (\ie, $\alpha\in \{2,20\}$) is preferred to balance the original and the reflected representations, while $\phi_4$ is likely to lead to the failure on larger $\alpha$ and cannot be adaptive to input $\textbf{R}_{j}$. Such results indeed verify the effectiveness of our adaptive decay function in \emph{\model}. The models using $\phi_1$ and $\phi_2$ are the best ones and they show very close prediction performance.

\begin{figure}
    \centering
    \subfigure[][Hidden Dim. Test on ML-1M]{
        \centering
        \includegraphics[width=0.32\textwidth]{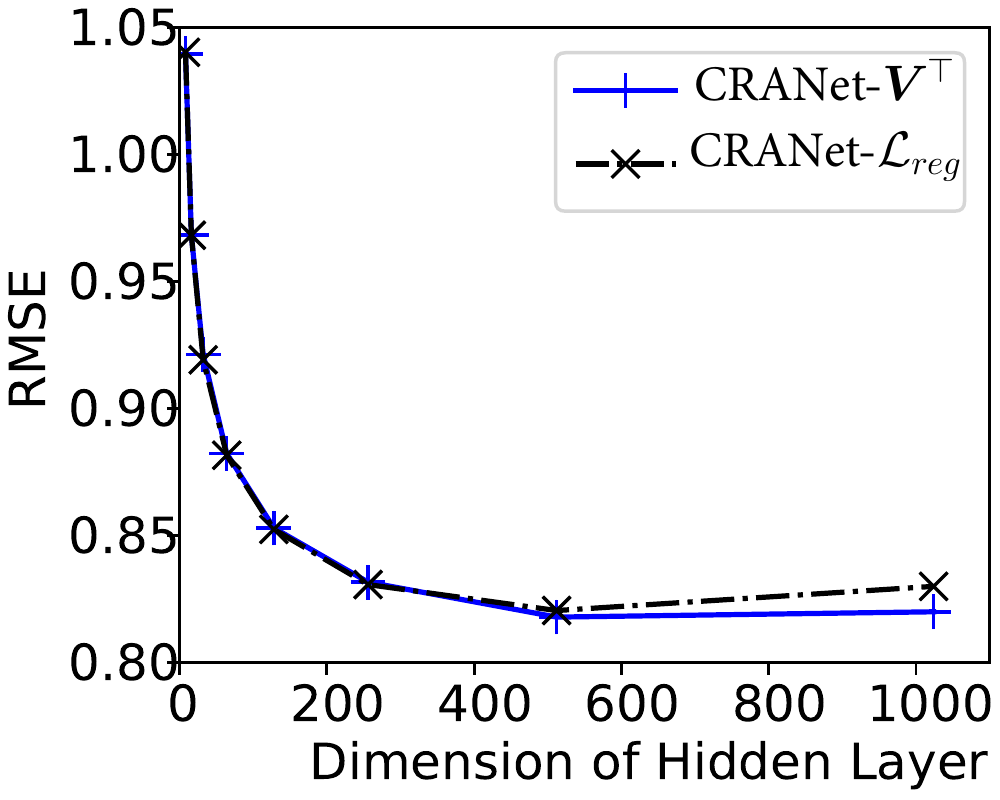}
        \label{fig:hid_ml1m}
        }
    \subfigure[][Hidden Dim. Test on ML-10M]{
        \centering
        \includegraphics[width=0.32\textwidth]{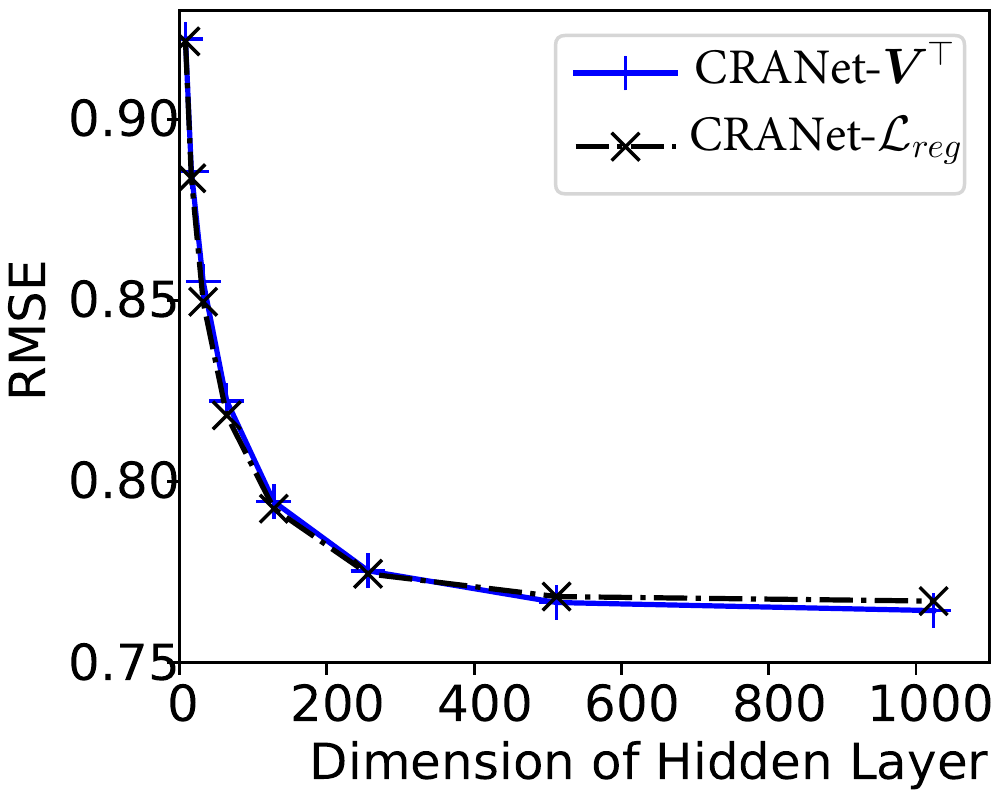}
        \label{fig:hid_ml10m}
        }
	\caption{Model Performance v.s. hidden layer dimensionality on Movielens-1M and Movielens-10M Data}
    \label{fig:intTest}
    \vspace{-0.1in}
\end{figure}

\begin{table}
		\centering
		\footnotesize
		\caption{\footnotesize{Effect of the Reflection Reception.}}
		\label{tab:reflection_reception_ablation}
		\vspace{-0.05in}
		\begin{tabular}{cccc}
        	\toprule
        	Model & ML-1M & ML-10M & Netflix\\
        	\midrule	\emph{\model}-R&0.831&0.782&0.823\\	\emph{\model}-N&0.838&0.779&0.821\\	\emph{\model}&\textbf{0.822}&\textbf{0.768}&\textbf{0.814}\\
        	\bottomrule
\end{tabular}
\end{table}

\begin{table}
		\centering
		\footnotesize
		\caption{\footnotesize{Best performance of \emph{\model} for different decay functions with the corresponding $\alpha$.}}
		\label{tab:best_perform_decay}
		\vspace{-0.05in}
		\begin{tabular}{cccccccccccc}
            \toprule
            \multirow{2}{*}{Dataset} & \multicolumn{2}{c}{$\phi_1$} && \multicolumn{2}{c}{$\phi_2$} && \multicolumn{2}{c}{$\phi_3$} && \multicolumn{2}{c}{$\phi_4$}\\
            \cline{2-3}\cline{5-6}\cline{8-9}\cline{11-12}
            & $\alpha$ & RMSE && $\alpha$ & RMSE && $\alpha$ & RMSE && $\alpha$ & RMSE\\
            \midrule
            ML-1M & 20 & \textbf{0.822} && 2 & \textbf{0.822} && 2 & 0.823 && 0.02 & 0.827\\
            ML-10M & 20 & \textbf{0.768} && 2 & 0.769 && 0.2 & 0.770 &&  0.02 & 0.773\\
            \bottomrule
        \end{tabular}
	\vspace{-0.05in}
\end{table}


\subsection{Performance \emph{v.s.} Data Sparsity (RQ5)}
\begin{figure}[h!]
    \centering
    \vspace{-0.1 in}
    \subfigure[][\scriptsize{ML-1M Item-Based}]{
        \centering
        \includegraphics[width=0.25\textwidth]{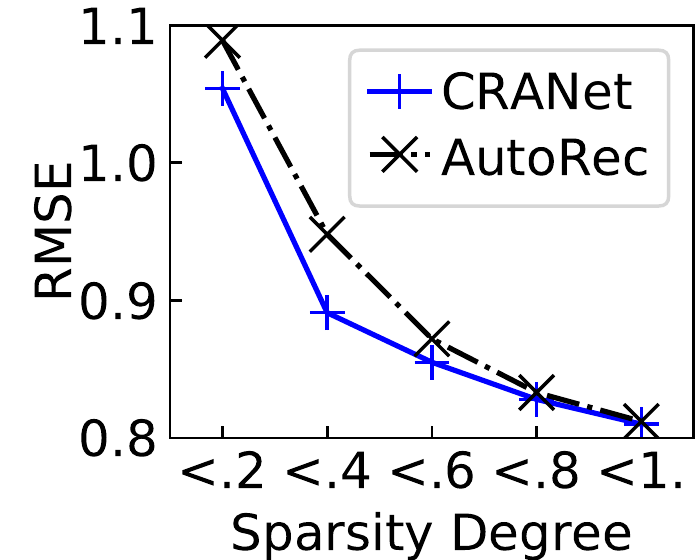}
        \label{fig:I_ML_1M}
        }
    \subfigure[][\scriptsize{ML-10M Item-Based}]{
        \centering
        \includegraphics[width=0.25\textwidth]{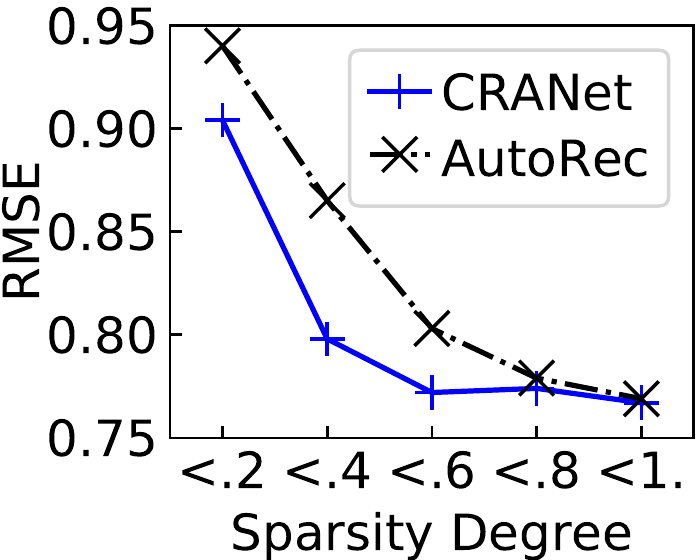}
        \label{fig:I_ML_10M}
        }
    \subfigure[][\scriptsize{Netflix Item-Based}]{
        \centering
        \includegraphics[width=0.25\textwidth]{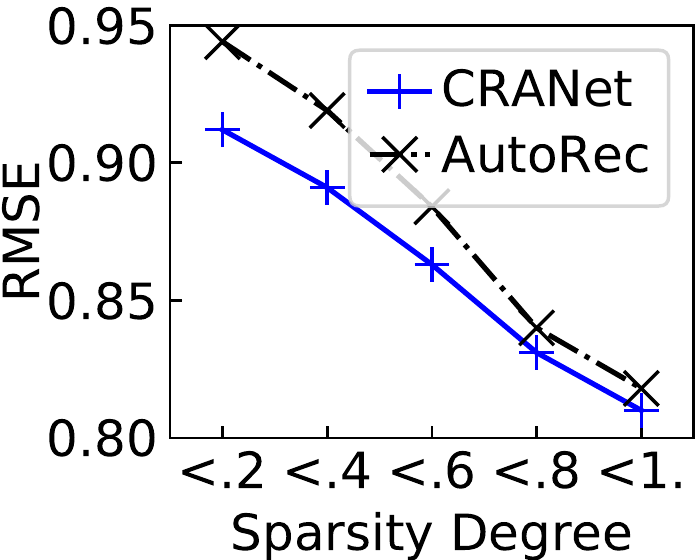}
        \label{fig:I_Netflix}
        }
    \subfigure[][\scriptsize{ML-1M User-Based}]{
        \centering
        \includegraphics[width=0.25\textwidth]{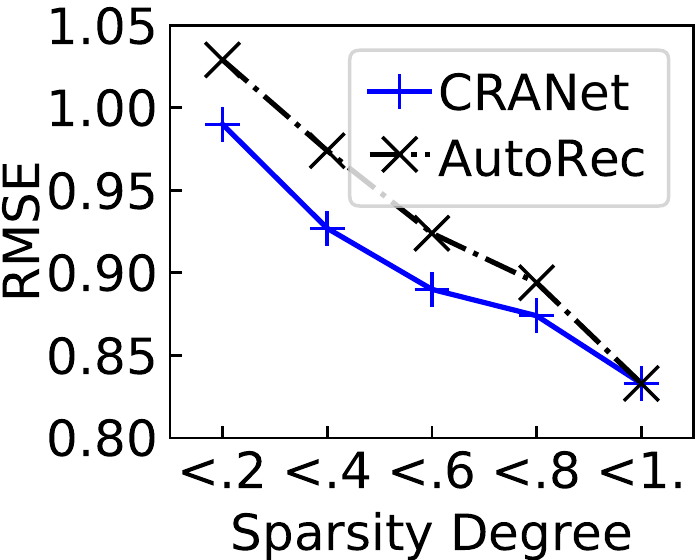}
        \label{fig:U_ML_1M}
        }
    \subfigure[][\scriptsize{ML-10M User-Based}]{
        \centering
        \includegraphics[width=0.25\textwidth]{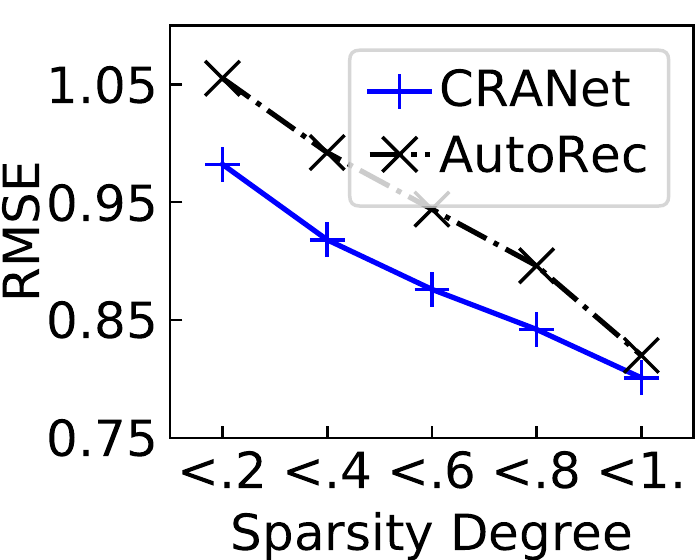}
        \label{fig:U_ML_10M}
        }
    \subfigure[][\scriptsize{Netflix User-Based}]{
        \centering
        \includegraphics[width=0.25\textwidth]{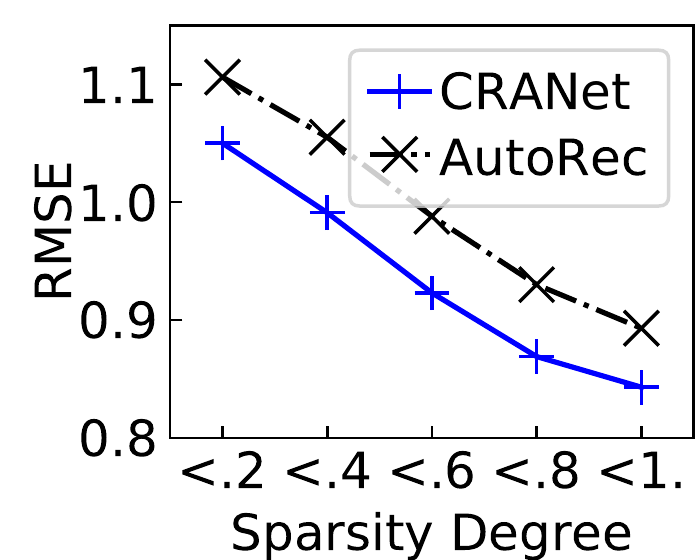}
        \label{fig:U_Netflix}
        }
    \vspace{-0.15in}
    \caption{Performance evaluation with varying sparsity degree. The $x$-axis denotes the interval of sparsity degree on the sub-dataset, \eg, the left-most two points are the test results on the sparsest 20\% input data. The $y$-axis denotes the average RMSE scores.}
    \label{fig:testlines}
\end{figure}

We perform experiments by varying sparsity degrees of input user-item interaction data, to investigate the capability of \emph{\model} in dealing with sparse inputs. In specific, each dataset containing items' rating vectors is divided into five subsets equally, with the first sub-dataset containing the sparsest $20\%$ input vectors and their corresponding test vectors, the second one containing the second sparsest $20\%$ vectors, and so on. The model training and test process are separately performed for each dataset. The experimental results of the proposed \emph{\model} and AutoRec are shown in Fig.~\ref{fig:testlines}. We could notice that \emph{\model} consistently outperforms AutoRec in most evaluation cases, which further demonstrates that the reflection mechanism is very helpful in modeling unobserved user-item relations from highly sparse user-item interaction data.

\begin{table}[h!]
	\small
	\caption{Performance of \model\ \wrt\ Different Reflection Mechanisms.}
	\vspace{-0.05in}
	\label{tab:userVsItem}
	\centering
	\setlength{\tabcolsep}{1.5mm}
	\begin{tabular}{ccccc}
		\toprule
		Item / User & Reflection Mechanism & ML-1M & ML-10M & Netflix\\
		\midrule
		\multirow{2}*{User-based}
		&$\textbf{V}\tp$&{0.858}&{0.813}&{0.849}\\
		&$\textbf{T}$ with $\mathcal{L}_\text{reg}$&0.860&0.831&0.868\\
		\hdashline
		\specialrule{0em}{1pt}{1pt}
		\multirow{2}*{Item-based} &$\textbf{V}\tp$&\textbf{0.822}&0.769&0.818\\
		&$\textbf{T}$ with $\mathcal{L}_\text{reg}$&\textbf{0.822}&\textbf{0.768}&\textbf{0.814}\\
        \hline
	\end{tabular}
	\vspace{-0.05in}
\end{table}

\subsection{Item-based \emph{v.s.} User-based \emph{\model} Performance Investigation (RQ6)}
To explore the impact of item-based and user-based recommendation framework, it is interesting to see how our \emph{\model} performs when generating the user-item interaction matrix $\textbf{R}$ from either user or item side. Table~\ref{tab:userVsItem} lists the evaluation results of \emph{\model} with the designed different reflection mechanism (\ie, transpose-based reflection gate--$\textbf{V}\tp$ and implicit reflection gate--$\textbf{T}$ with $\mathcal{L}_\text{reg}$). We can observe that the item-based \emph{\model} framework consistently outperforms the user-based \emph{\model} framework with different network configurations, which is consistent with the observations of state-of-the-art autoencoder-based recommendation techniques~\cite{ouyang2014autoencoder, sedhain2015autorec, strub2016hybrid}. The performance gap may result from the lower density degrees of user's interaction vectors as compared to that of items, which leads to the overfitting.

\begin{table*}
	\footnotesize
	\vspace{-0.05in}
	\caption{Model scalability study with computational cost (measured by running time seconds) and the corresponding model size (\# of trainable parameters in millions).}
	\vspace{-0.05in}
	\label{tab:Exonold}
	\begin{tabular}{ c | c | c | c | c | c | c }
		\toprule
		Dataset  & \multicolumn{2}{|c|}{ML-1M} & \multicolumn{2}{|c|}{ML-10M} & \multicolumn{2}{|c}{Foursquare} \\ 
		\hline
		Method   & Time & Params \# & Time & Params \# & Time & Params \# \\
		\hline
		BiasMF   & 1.80s & 4.89M & 19.03s & 40.44M   & 20.75s & 16.29M \\
		DMF      & 5.18s & 4.87M & 57.24s & 40.28M   & 49.63s & 16.26M \\
		NCF      & 1.96s & 5.37M & 27.29s & 40.78M   & 21.43s & 16.76M \\
		ACF      & 1.53s & 30.23M & 27.45s & 699.48  & 25.68s & 27.19M \\
		AutoRec  & 1.00s & 6.05M & 19.15s & 69.95M   & 18.94s & 7.77M  \\
		CDAE     & 1.41s & 6.05M & 19.54s & 69.95M   & 19.08s & 7.77M  \\
		NADE     & 3.61s & 30.23M & 39.61s & 10.83M  & 25.14s & 27.19M \\
		CF-UIcA  & 2.37s & 48.78M & 29.05s & 806.36M & 26.94s & 113.85M\\
		NGCF     & 4.52s & 24.87M & 41.90s & 403.28M & 44.08s & 81.78M\\
		ST-GCN   & 3.97s & 24.62M & 32.40s & 403.03M & 37.96s & 81.53M\\
		\hline
		\emph{\model} & 1.09s & 6.30M & 18.54s & 70.72M & 19.01s & 8.02M\\
		\hline
	\end{tabular}
	\vspace{-0.1in}
	\label{tab:scalability}
\end{table*}


\subsection{Scalability Study of \model\ (RQ7)}
\label{sec:model_efficiency}
In addition to evaluating the model effectiveness, we also investigate the efficiency of \emph{\model} and state-of-the-art baselines as reported in Table~\ref{tab:scalability} (experiments are performed on one Titan V card). We can observe that the computational cost (\ie, measured by running time and the corresponding model size) of \emph{\model} is much less than that of NADE, which demonstrates the efficiency of \emph{\model}, \ie, our reflective reception network with implicit reflection mechanism is a scalable learning module in terms of the space and computational cost.

\begin{figure*}
    \includegraphics[width=0.96\textwidth]{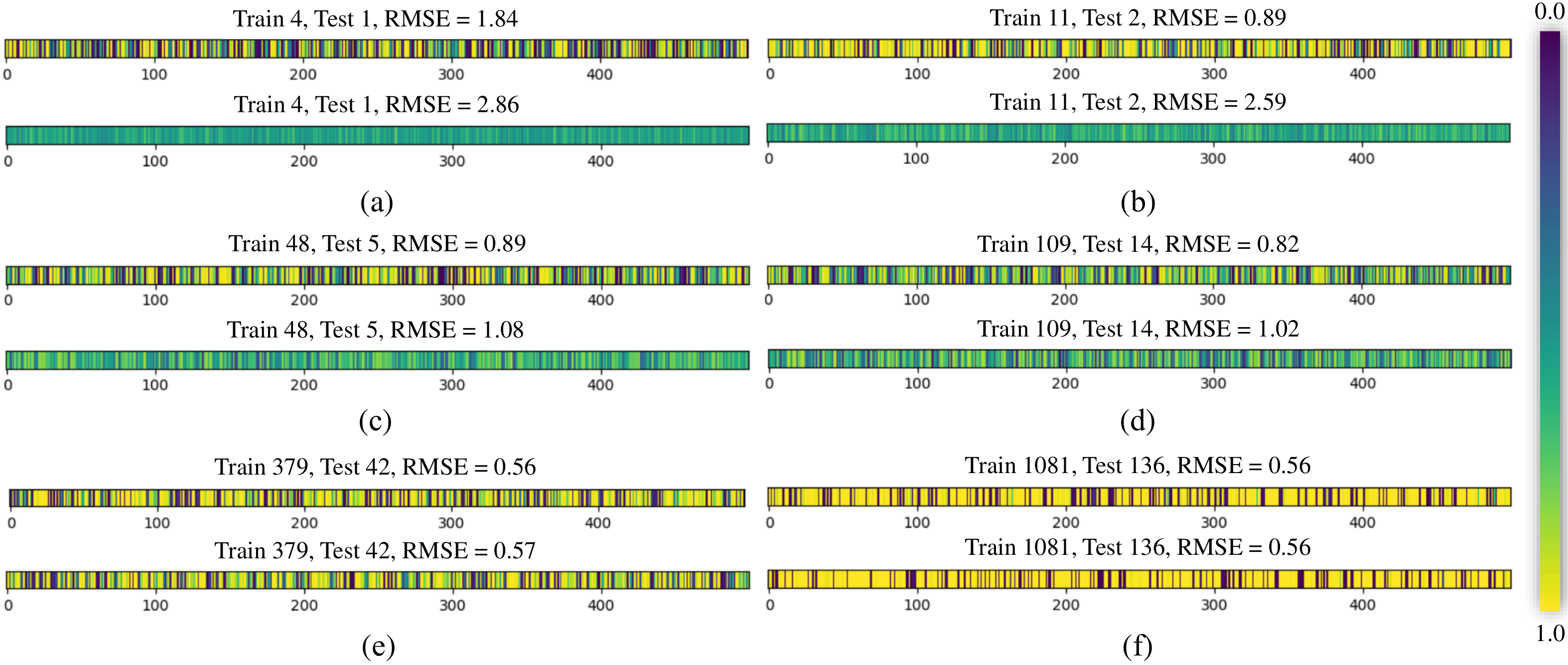}
    \caption{Typical samples of activated latent representations given by \model\ on ML-1M dataset. The samples are visualized by heat maps. The upper and bottom heatmaps in (a)-(f) show the representations for one data instance given by \model\ and AutoRec respectively, and each strip contains $500$ elements whose values are represented by colors. The relationship between color and value is shown by the color-bar in the legend: the brighter the color is, the larger the value will be. The numbers of observed ratings in training vectors, the numbers of corresponding tested ratings, and the prediction errors are also shown.}\label{fig:heatMap}
	\vspace{-0.05in}
\end{figure*}

\subsection{Hyperparameter Study of \emph{\model} (RQ8)}

We also explore the influence of hidden state dimensionality for the recommendation performance. During the parameter sensitivity study, we vary the target hyperparameter within a specific value ranges and keep other hyperparameters as default settings. The results of the item-based \emph{\model} framework with both the tied-weight and the implicit reflection schemes are shown in Figure~\ref{fig:hid_ml1m} and~\ref{fig:hid_ml10m}. The experimental results of two versions are quite similar. Furthermore, with the increasing of the dimension of hidden layer, the prediction error first sharply decreases, then varies little and finally increase a bit due to overfitting. According to the experimental results, a hidden layer containing around $500$ units causes no significant over-fitting or under-fitting. Models in other experiments use 500 as the dimensionality of hidden layers.

\subsection{Case Study}
We visualize the latent representations learned by \emph{\model} to investigate the interpretation ability of our reflective reception network. We show the representations of the corresponding hidden layer from several typical data instances and make comparison with the best performed autoencoder-based model variant \emph{\model}-R as shown in Fig.~\ref{fig:heatMap}. For data instances with few observations like Fig.~\ref{fig:heatMap}(a) and Fig.~\ref{fig:heatMap}(b), \emph{\model}-R generally gives representations which contain a lot of neutral values (close to 0.5, colored in green), due to the trade-off between the sparse and dense vectors. Such representations can hardly show good interpretation ability in encoding interactive relations between users and items.

In contrast, owing to the reflected representations normalized by data sparsity in \emph{\model}, we give such data instances almost full of activated (bright yellow) and suppressed (dark purple) elements, which is more distinguishable. Such higher discrimination brings the better performance of \emph{\model} in dealing with highly sparse vectors. We also found that both methods sometimes assign non-distinguishable representations to those with fairly enough observations (\eg~Fig.~\ref{fig:heatMap}(c) and Fig.~\ref{fig:heatMap}(d). This may be caued by the inherent characteristics of the data instances. In comparison, \emph{\model} can still give much more distinguishable representations, with more neutral values than those in Fig.~\ref{fig:heatMap}(a) and Fig.~\ref{fig:heatMap}(b).

%% file: conclusion.tex
\section{Conclusion}
\label{sec:conclusion}

This paper proposes a new reflective collaborative filtering framework (namely \model) to model implicit multi-modal user-item relations for better recommendation. The \model\ model integrates the designed reflection mechanism with a cross-modal autoencoder network for latent representation mapping. The newly developed reflection mechanism endows the collaborative filtering paradigm with the ability of encoding more informative user-item relations in correcting variable sparsity bias. Our proposed frameworks achieve significantly better performances as compared to state-of-the-art techniques on both rating prediction and item recommendation applications across multiple datasets. Since \model\ is a generic reflective collaborative filtering, our future work plans to extend \model\ with other recommendation architecture, such as factorization machines.
